\documentclass[11pt]{article}

\usepackage[a4paper]{geometry}

\usepackage{amsthm,amsmath,amssymb,amsfonts,amscd}
\usepackage{graphicx}
\usepackage{enumerate}
\usepackage[all]{xy}
\usepackage{pgf}
\usepackage{tikz}
\usetikzlibrary{arrows}
\usepackage{float}
\usepackage{hyperref} 

\usepackage[medium]{titlesec}
\titleformat{\section}%
{\fontfamily{lmss}\Large\bfseries}{\thesection}{1ex}{}
\titleformat{\subsection}%
{\fontfamily{lmss}\large\bfseries}{\thesubsection}{1ex}{}
\titleformat{\paragraph}[runin]%
{\fontfamily{lmss}\normalsize\bfseries}{\theparagraph}{1em}{}

\newtheorem{theorem}{Theorem}[section]

\newtheorem{lemma}[theorem]{Lemma}

\theoremstyle{definition}

\theoremstyle{remark}

\renewcommand{\d}{\mathrm{d}} 

\newcommand{\cp}[1]{\mathcal{C}^{#1}}

\def\calF{\mathcal{F}}

\begin{document}


\thispagestyle{empty}

\null
\bigskip

\begin{center}
\sffamily
\LARGE
The wave equation for stiff strings and piano tuning

\vskip 10mm

\Large
Xavier Gr\`acia%
\footnote{\textsf{email: xavier.gracia@upc.edu}}
and
Tom\'as Sanz-Perela%
\footnote{\textsf{email: tomassanzperela@gmail.com}}

\bigskip

Departament de Matem\`atiques
\\
Universitat Polit\`ecnica de Catalunya

\vskip 10mm

4 February 2016 /
revised 16 June 2016

\vskip 10mm
accepted for publication in
\textsl{Reports@SCM}

\end{center}

\bigskip

\begin{quote}
\parindent 0pt
\sffamily

\textbf{Abstract}

We study the wave equation for a string with stiffness.
We solve the equation and provide a uniqueness theorem
with suitable boundary conditions.
For a pinned string we compute the spectrum, 
which is slightly inharmonic.
Therefore, 
the widespread scale of 12 equal divisions of the just octave
is not the best choice 
to tune instruments like the piano.
Basing on the theory of dissonance,
we provide a way to tune the piano 
in order to improve its consonance.
A good solution is obtained 
by tuning a note and its fifth by minimizing their beats.




\end{quote}

\bigskip

\def\baselinestretch{1.05}
\normalsize

\section{Introduction}

\label{sec:intro}


The problem of finding appropriate scales
for playing music 
has worried musical theorists and instrument makers
for centuries. 
There is a close relationship 
between the theory of musical scales 
and
the frequency spectra of musical instruments;
indeed, 
the harmonic spectrum of most instruments
has lead to the present-day tempered scale,
with 12 \emph{equal divisions of the octave}
(12-edo).
However, piano strings have some degree of stiffness,
which implies that their spectrum is slightly inharmonic,
and this explains why the tuning of the piano is actually
``stretched'', 
with octaves slightly larger than should
\cite[p.\,389]{FletcherRossing}.
The purpose of this paper is 
to perform an accurate mathematical study 
of the wave equation of a string with stiffness,
and what does it imply to the choice of a scale.
Throughout the paper we will assume an elementary knowledge
of acoustics (physical and perceptive properties of the sound)  
and
of music theory (intervals and the tempered scale);
for the benefit of the reader, 
we have collected some of these notions in an appendix.

The classical wave equation 
($u_{tt} = c^2 u_{xx}$) 
models a perfectly elastic string. 
If we want to take the stiffness into account
we need to modify the equation. 
The simplest way to do this
consists in adding a term coming from the Euler--Bernoulli beam equation,
which is used to model the deflection of rigid bars. 
The result is a fourth order PDE of the form
$
u_{tt} = c^2 u_{xx} - M^2 u_{xxxx}
$
that has been seldom studied in the acoustics literature
\cite{FletcherRossing,MorseIngard},
and often sketchily.
This is why we have found it convenient 
to perform a more detailed and self-contained study 
with rather elementary techniques
(Section~2).
We compute 
the explicit form of the solutions, which turns out to be
the same as in the non-stiff case, 
except by the fact that the frequency spectrum
is no longer harmonic, but of the form
$f_n = n\,f_\circ\,\sqrt{1+Bn^2}$ 
($n \geq 1$),
where $B$ is a constant depending on the physical parameters of the string.
We also show the existence and uniqueness for the PDE
with appropriate boundary conditions.

For piano strings the value of the inharmonicity parameter $B$ is about
$10^{-3}$.
This means that 
its spectral frequencies slightly deviate from the harmonic ones
(it has a ``stretched harmonic spectrum'').
Though small, 
this deviation is of great importance 
for the consonance of 
the intervals between notes,
because the human ear is very sensitive to frequency differences.

The auditory perception qualifies some musical sounds 
as \emph{consonant} (``pleasant''), 
whereas others are \emph{dissonant}.
As it is explained in more detail in Section~3 and the Appendix, 
there is a close relationship between dissonance, spectrum and scale:
the choice of the notes used to play music
aims to achieve the best possible consonance,
and this consonance depends directly on the spectrum of the sounds.
Therefore, 
the fact that stiff strings have a slightly inharmonic spectrum
leads to reconsider 
the exact tuning of the notes we play with them.
A tool to perform a systematic study of this problem
is the \emph{dissonance curve} of a spectrum.
Basing on experimental results by Plomp and Levelt 
\cite{PlompLevelt},
one can define a function to measure the dissonance of two notes
as a function of their frequency ratio,
and draw a dissonance curve,
which depends strongly on the spectrum.
The local minima of this curve indicate possible good choices for
the notes of the scale
as far as the consonance of its intervals is concerned
\cite{Sethares}.

We apply this approach to the string with stiffness.
Its frequency spectrum is 
given by
$f_n = n\,f_\circ\,\sqrt{1+Bn^2}$
(see Section~2)
and therefore the ratio between the first and the second partials
is not the just octave 2:1,
but a ``stretched octave'' 
$2 \sqrt{1+4B} / \sqrt{1+B}$.
So we wonder if there exist scales
for this spectrum
that could possibly be more ``consonant''
than the usual 12-edo scale.
Aiming to preserve the freedom to modulate to any tonality,
we look for a scale with equal steps,
or, equivalently,
equal divisions of a certain interval, 
as for instance a stretched octave.
In Section~4 we use
the dissonance curve of the stretched spectrum
to study this problem in two different ways.
One is based in the coincidence of some partials.
The other one minimizes a weighted mean of the dissonance.
As a result, we obtain that
a good solution is to tune the fifth by 
making the second partial of the higher note
to coincide with the third partial of the fundamental note.

The paper is organised as follows.
In Section~2 we study the modelling of a string with stiffness:
we give an explicit solution of the equation
when the boundary conditions are those of a pinned string,
we present a rigorous derivation of its spectrum, 
and we state a uniqueness theorem.
In Section~3 we recall some facts about the theory of dissonance 
and how to draw dissonance curves,
and we obtain the dissonance curve of the string with stiffness.
In Section~4 we study several proposals to tune the piano,
either based in the coincidence of partials
or the minimization of the mean dissonance.
Section~5 is devoted to conclusions.
Finally, an appendix gathers some basic concepts of 
acoustics and music theory.

\section{The wave equation for the string with stiffness}

It is well known 
\cite{Salsa,Strauss} 
that the motion of a vibrating string (for instance, a
violin string) can be represented by the solutions of the problem
\begin{equation}
\label{eq: cauchyProblemIdeal}
\begin{cases}
\dfrac{\partial^2 u}{\partial t^2} = c^2\dfrac{\partial^2 u}{\partial x^2} &
x\in(0,L),\ t>0\\
u(0,t) = u(L,t) = 0 & t\geq0 \\
u(x,0) = \phi(x) & x\in[0,L] \\
\partial_t u (x,0)= \psi(x)& x\in[0,L]\, ,
\end{cases}
\end{equation}
where $u(x,t)$ represents 
the transversal displacement of the string of length $L$ 
(represented by the interval $[0,L]$) 
from its equilibrium position, and
$\phi(x)$ and $\psi(x)$ are, respectively, 
the initial shape and velocity of the string.
The boundary conditions 
$u(0,t) = u(L,t) = 0$
for
$t \geq 0$ 
mean that the string
has fixed ends and $c^2 = \tau/\rho$, 
with $\tau$ the tension of the string and $\rho$ its linear density.
The value $c$ is the velocity of the travelling waves along the string.

The solution of this equation can be computed using the method of separation of
variables, 
and we obtain
\vadjust{\kern -3mm} 
\begin{equation}
\label{eq: solutionIdeal}
u(x,t) = \sum_{n=1}^\infty \left \lbrack a_n\cos{\left ( 2\pi f_n
t\right )} + b_n\sin{\left ( 2\pi f_n t\right )}\right \rbrack
\sin{\left ( \frac{n\pi}{L} x\right )}\,, \quad f_n = \dfrac{n c}{2 L}
\end{equation}
where the coefficients $a_n$ and $b_n$ are obtained from 
the Fourier coefficients of the initial conditions $\phi$,~$\psi$.
For the convergence and smoothness of this series
some regularity conditions are required on $\phi$,~$\psi$,
see for instance
\cite{Salsa}.

The model of the wave equation is a good approximation for instruments like the
guitar, whose strings are almost perfectly flexible. 
However, when we want to
model the motion of the piano strings, which have greater stiffness, the classical
wave equation is not good enough. 
For this reason, 
a term describing the resistance against bending is added to it
(see 
\cite[p.\,64]{FletcherRossing}),
obtaining the following equation:
\begin{equation}
\label{eq: stiffwaveequation}
\dfrac{\partial^2 u}{\partial t^2} = c^2\dfrac{\partial^2 u}{\partial x^2}-
\dfrac{ESK^2}{\rho} \dfrac{\partial^4 u}{\partial x^4}\,,
\end{equation}
where $S$ is the cross-sectional area of the string, $E$ is Young's modulus of its material,
$\rho$ is its linear density and $K$ is the radius of gyration,
which is $K=R/2$
for a cylindrical shape of radius~$R$
\cite[p.\,58]{FletcherRossing}.

The added term is the same that appears in the beam equation 
(also called Euler--Bernoulli equation), 
that models the motion of a vibrating beam 
under the hypotheses of no shear stress nor rotational inertia;
a deduction of this equation can be found, for instance,
in
\cite[p.\,58]{FletcherRossing},
or at the end of Chapter~2 of
\cite{Tikhonov}.
One can view \eqref{eq: stiffwaveequation} 
as the generalization of a PDE for a vibrating material: 
the first term is due to the elasticity of the material 
(its capacity to return to the initial position after a
deformation) 
and the second one, due to the resistance against bending. 
If the 
first term is zero, the material is not elastic and we
get the beam equation. 
On the contrary, if the second term is zero, the material is not rigid and we obtain the wave equation.

The string with stiffness 
based on the Euler--Bernouilli model 
is the most widely used model.
Nevertheless, there exist other equations 
that can model the vibration of rigid materials 
(and in particular piano strings). 
Prominent among them is the Timoshenko beam model, 
which takes into account shear stress and rotational inertia
\cite{Han}.
The description of the motion of piano strings 
using this model has been thoroughly discussed recently in the thesis~%
\cite{Chabassier}.
It is shown there that the frequencies of 
the string based upon the Timoshenko beam model 
behave as the ones based on the Euler--Bernouilli model for the lower
partials;
the Timoshenko model provides a better description 
for higher partials,
a region where their contribution to dissonance is negligible.
Therefore the Euler--Bernouilli model is enough for our purposes.

\subsection{Solving the equation}

Equation
\eqref{eq: stiffwaveequation}
was studied in
\cite{Fletcher},
where the author guesses the form
of some solutions with separate variables.
Besides that article, 
only a few references in the acoustics literature
deal with the string with stiffness, 
and they merely give approximate solutions of the spectrum,
without further justification. 
A recent study of this equation is in 
\cite{Chabassier},
where the exact formula for the frequency of the partials 
is found using Fourier transform,
though the equation is not actually solved.
So we have found it convenient to perform a detailed study:
by following the standard method of separation of variables
we give a solution of the initial value problem 
with appropriate boundary conditions, obtaining 
also the formula for the frequencies.
Uniqueness of the solution is studied in the following section.

We start by looking for a solution 
to equation~\eqref{eq: stiffwaveequation}
of the form
$u(x,t) = X(x)T(t)$.
We have
\begin{equation}
\label{eq: separationVariables}
X T'' = c^2 X'' T -  \dfrac{ESK^2}{\rho} X^{(4)}T
\quad \Longrightarrow \quad
\dfrac{T''}{T} =
c^2\dfrac{X''}{X}- \dfrac{ESK^2}{\rho} \dfrac{X^{(4)}}{X} 
\,.
\end{equation}
As the left-hand side of the equation depends only on $t$ and the right-hand
side, only on $x$, \eqref{eq: separationVariables} has to be a non
positive constant, called $-\omega^2$ (non
positive because we are looking for periodic solutions in time): 
\begin{equation}
\label{eq: separationVariables2} 
\dfrac{T''}{T} 
= 
c^2\dfrac{X''}{X}- \dfrac{ESK^2}{\rho} \dfrac{X^{(4)}}{X} 
= 
-\omega^2 .
\end{equation}
If we look at the time equation, we have an ODE which is easy to solve:
$T_\omega(t) = A \cos{\omega t} + B\sin{\omega t}\,$.


We look now for the solutions of the ODE for $X$:
\begin{equation}
\label{eq: spatialODE}
\dfrac{ESK^2}{\rho} X^{(4)} - c^2 X'' - \omega^2 X = 0\,.
\end{equation}
We divide the equation by $ESK^2 / \rho$ and define
$a := c^2 \rho/ESK^2$ and 
$b := \rho \omega^2/ESK^2$.
After that, \eqref{eq: spatialODE} becomes 
$X^{(4)} - a X'' - b X = 0\,,$ 
whose solutions are of the form
\begin{equation}
\label{eq: spatialODEgeneralSoluion}
C_1 \cosh{k_1 x} + C_2 \sinh{k_1 x}+ C_3 \cos{k_2 x} + C_4 \sin{k_2 x}\,,
\end{equation}
with
$
k_1  = \sqrt{\dfrac{a + \sqrt{a^2 + 4b}}{2}}
$
and
$
k_2  = \sqrt{ \dfrac{-a + \sqrt{a^2 + 4b}}{2}}\,.
$
We introduce again two convenient constants:
\begin{equation}
\label{eq: def B f}
B := \pi^2 \dfrac{ESK^2}{\tau L^2} 
\quad \text{and} \quad 
f_\circ := \dfrac{c}{2L}\,.
\end{equation}
In this way, using the definition of $a$ and $b$, 
we obtain the following relations between
$k_1, k_2$ and~$\omega$:
\begin{equation}
\label{eq: def k1 k2}
k_1^2 = 
\dfrac{\pi^2}{2BL^2}
\left[ \sqrt{1 + \dfrac{\omega^2 B}{f_\circ^2 \pi^2}} + 1 \right]
\quad \text{and} \quad 
k_2^2 = 
\dfrac{\pi^2}{2BL^2}
\left[ \sqrt{1 + \dfrac{\omega^2 B}{f_\circ^2 \pi^2}} - 1 \right] .
\end{equation}

We want to find the possible values of $k_1$ and $k_2$, 
that will determine the
possible values of $\omega$. 
In order to do it, we will impose the boundary
conditions, but now, as the equation is of 4th order, 
we need 4 boundary conditions, two more
apart from the Dirichlet boundary conditions on both ends of the string.
We will consider two cases:

$\bullet$ $X' = 0$ at the ends.
This case appears when the string is clamped at the ends.

$\bullet$ $X'' = 0$ at the ends.
This happens when the string is pinned at the ends,
since there is no moment.

The first case, 
$X = X' = 0$ at the ends of the string, 
leads to an equation that can be solved numerically, 
but is not possible to get a closed formula 
for the spectrum of frequencies 
(see \cite{Fletcher} for more details and for an approximate formula). 
The second case, 
$X = X'' = 0$ at the ends of the string, 
is easier to solve and 
will lead us to a formula for the frequencies of the partials. 
In the case of the piano, 
this second option seems to be closer to reality, 
because the strings are supported on a bridge.
From now on, we will focus in this case.

\paragraph{Pinned boundary conditions}

We are going to solve the problem with the condition 
$X = X'' = 0$ at the ends of the string.
Consider a general solution of \eqref{eq: spatialODE}
\begin{equation}
\label{eq: generalSolution 3}
X(x) = C_1 \cosh{k_1 x} + C_2 \sinh{k_1 x} + C_3 \cos{k_2 x} + C_4 \sin{k_2 x}.
\end{equation}
We want to find the possible values of $k_1$ and $k_2$ that make  
\eqref{eq: generalSolution 3} 
satisfy (non-trivially) the boundary conditions.
Let us 
impose the boundary conditions at the string ends, $x=0,L$.

For $x=0$, we obtain:
\begin{equation}
X(0) = C_1 + C_3 = 0 \qquad\text{and} \qquad 
X''(0) = C_1 k_1^2 - C_3 k_2^2 = 0\,.
\end{equation}
From the first equation we get $-C_3=C_1$ and, 
replacing it in the second one,
we arrive to the equation $C_1 (k_1^2 + k_2^2 ) = 0$, 
which implies $C_1 = C_3 = 0$.

Now we impose the boundary conditions at $x=L$ to $X(x) =
C_2\sinh{k_1 x} + C_4\sin{k_2 x}$, obtaining:
\begin{equation}
X(L) = C_2\sinh{k_1 L} + C_4\sin{k_2 L} = 0 \qquad\text{and} \qquad 
X''(L) = C_2 k_1^2\sinh{k_1 L} - C_4 k_2^2\sin{k_2 L} = 0.
\end{equation}
Multiplying the first equation by $k_2^2$ and adding it to the second one, we
get $C_2 (k_1^2 +  k_2^2 ) \sinh{k_1 L} = 0$. 
As the last two factors are different from zero, 
we conclude that $C_2=0$.

Finally, we have $C_4 \sin{k_2 L} = 0 $. 
As we want nontrivial solutions, we need
$C_4 \neq 0$ and, thus, $k_2 L = n \pi$ for $n = 1, 2, \ldots$. 
From this relation and \eqref{eq: def k1 k2}
we obtain
\begin{equation}
\left( \dfrac{n \pi}{L} \right) ^2 = 
\dfrac{\pi^2}{2BL^2}\left[ \sqrt{1 + \dfrac{\omega_n^2 B}{f_\circ^2 \pi^2}} - 1 \right]
\end{equation}
and, isolating $\omega_n$, we obtain the possible frequencies:
\begin{equation}
\label{eq: freq stiffness}
f_n = \dfrac{\omega_n}{2 \pi} = n \, f_\circ \sqrt{1 + B n^ 2}
\quad \text{with} \quad 
n=1,2,\ldots
\end{equation}
Thus, for each of the~$\omega_n$,
the solution of \eqref{eq: spatialODE} 
that satisfies the boundary conditions is a multiple of
\begin{equation}
\label{eq: spatialODE Xn solutions}
X_n(x) = \sin{\left(\dfrac{n \pi}{L} x\right)} 
\quad \text{with} \quad 
n = 1, 2, \ldots
\end{equation}
Remarkably, 
these are the same modes of vibration as in the case without
stiffness: the difference only shows up in the frequencies of vibration.

To conclude, we can write the general solution of the PDE 
\eqref{eq: stiffwaveequation}, 
with boundary conditions 
$u = 0$ and $u_{xx} = 0$ at the ends of the string, as:
\begin{equation}
\label{eq: general solution pinned}
u(x,t) = 
\sum_{n=1}^\infty 
\Big[
  a_n\cos{\left (2 \pi f_n t\right )} + 
  b_n\sin{\left (2 \pi f_n t\right )}
\Big] 
\sin {\left(\dfrac{n \pi}{L} x\right)}
\,,
\qquad
f_n = n \, f_\circ \sqrt{1 + B n^ 2}
\,,
\end{equation}
where $a_n, b_n$ are obtained from initial conditions in the same way as in the
ideal case. 

As we can see, 
the spectrum is no longer harmonic, but it is `stretched' from 
the harmonic one due to the factor $\sqrt{1 + B n^ 2}$. 
For a cylindrical string of radius~$R$ the value of~$B$ is 
$B = \dfrac{\pi^3}{4} \dfrac{ER^4}{\tau L^2}$;
its typical values for a piano string are about $10^{-3}$.

Notice that
the constant $f_\circ = c/2L$ would be the fundamental frequency of the string 
if it did not have stiffnes ($B=0$);
in this case
we would recover the frequency spectrum of the ideal string.
When $B > 0$, 
the fundamental frequency is $f_1 = f_\circ\sqrt{1+B}$, 
higher than~$f_\circ$.

\subsection{Uniqueness of solutions}

We prove now a theorem of uniqueness of solutions 
for the wave equation with
stiffness. 
This can be seen a particular case of the results of 
semigroup theory for evolution problems with monotone operators 
(see 
\cite{Brezis,Evans}), 
but in this case we provide an elementary proof,
similar to the uniqueness theorem for the wave equation, 
which can be found for instance in
\cite{Salsa}.
We will use the notation 
$\partial^n_{\xi} :=
\dfrac{\partial^n}{\partial \xi^n}$ 
when necessary.
\begin{lemma}
\label{lemma: energy}
Let $u(x, t) \in \cp{4}([0,L]\times[0, \infty))$ satisfying
\begin{equation}
\label{eq: stiffwaveequationS}
\dfrac{\partial^2 u}{\partial t^2} 
= 
c^2\dfrac{\partial^2 u}{\partial x^2} -
M^2 \dfrac{\partial^4 u}{\partial x^4}
\,, 
\quad 
x\in(0,L),\ t>0
\,.
\end{equation}
In any of the two following cases
\begin{equation}
\begin{cases}
u(0,t) = u(L,t)=0 & t\geq0 \\
\partial_x u(0,t) = \partial_x u(L,t)=0 & t\geq0 
\end{cases} \
\text{ \ or \ } \
\begin{cases}
u(0,t) = u(L,t)=0 & t\geq0 \\
\partial^2_x u(0,t) = \partial^2_x u(L,t)=0 & t\geq0 
\end{cases}
\:,
\end{equation}
the quantity
\begin{equation}
\mathcal{E}(u) = 
\dfrac{1}{2} \int_0^L 
\bigg(
(\partial_t u)^2 + c^2(\partial_x u)^2 +
M^2(\partial_x^2 u)^2 
\bigg)
\,\d x
\end{equation}
is constant in time.

\begin{proof}
We just need to show that the derivative with respect to $t$ of
$\mathcal{E}(u)$ is zero.
\begin{align*}
\dfrac{\d}{\d t} (\mathcal{E}(u)) 
& = 
\int_0^L 
\bigg(
\partial_t u \, \partial_t^2 u+ c^2
\partial_x u  \, \partial_t \partial_x u + M^2 \partial_x^2 u  \,\partial_t
\partial_x^2 u 
\bigg)
\d x 
\qquad\qquad\qquad\qquad\qquad\
\text{(using \eqref{eq: stiffwaveequationS})}
\\
& = 
\int_0^L 
\bigg(
\partial_t u (c^2 \partial_x^2 u - M^2 \partial_x^4 u) + c^2
\partial_x u \,\partial_t \partial_x u + M^2 \partial_x^2 u \,\partial_t
\partial_x^2 u 
\bigg) 
\d x
\\
& = 
c^2 \! \int_0^L \!\!
\bigg(
\partial_t u \, \partial_x^2 u + \partial_x u \,\partial_x \partial_t u 
\bigg)
\d x 
+ 
M^2 \!\! \int_0^L \!\!
\bigg(
\partial_x^2 u \, \partial_x^2 \partial_t u - \partial_t u \, \partial_x^4 u 
\bigg)
\d x 
=: 
c^2 I_c + M^2 I_S
\,.
\end{align*}

Now,
\begin{align*}
I_c & 
= 
\int_0^L 
\bigg(
\partial_t u \, \partial_x^2 u + \partial_x u \, \partial_x \partial_t u 
\bigg)
\d x  
\overset{\underset{\mathrm{parts}}{}}{=}  
\int_0^L 
\bigg(
\partial_t u \, \partial_x^2 u - \partial_x^2 u \, \partial_t u 
\bigg)
\d x  
+
\bigl[\partial_x u  \, \partial_t u \bigr]_0^L 
= 
\bigl[\partial_x u  \, \partial_t u \bigr]_0^L \,,
\end{align*}
which is zero due to the Dirichlet boundary condition 
($u = 0$ at $0,L$ for all
$t\geq0$ implies $\partial_t u = 0 $ for all
$t\geq0$).

Similarly, we have
\begin{align*}
I_S & 
= 
\int_0^L 
\bigg(
\partial_x^2 u  \: \partial_x^2 \partial_t u -
\partial_t u  \: \partial_x^4 u 
\bigg)
\d x  
\overset{\underset{\mathrm{parts}}{}}{=} 
\int_0^L 
\bigg( 
- \partial_x^3 u \:  \partial_x \partial_t u -
\partial_t u  \: \partial_x^4 u 
\bigg) 
\d x 
+
\bigl[\partial_x^2 u \: \partial_x\partial_t u \bigr]_0^L 
\\
& \overset{\underset{\mathrm{parts}}{}}{=}  
\int_0^L 
\bigg(
\partial_x^4 u \: \partial_t u -
\partial_t u   \: \partial_x^4 u 
\bigg) 
\d x 
+
\bigl[\partial_x^2 u  \: \partial_x\partial_t u \bigr]_0^L 
- \bigl[\partial_x^3 u \: \partial_t u \bigr]_0^L 
=
\bigl[\partial_x^2 u  \: \partial_x\partial_t u \bigr]_0^L 
- \bigl[\partial_x^3 u \: \partial_t u \bigr]_0^L  \,,
\end{align*}
which is again zero due to the boundary conditions. 

Therefore, we get $ \dfrac{\d}{\d t} (\mathcal{E}(u)) = 0\,.$
\end{proof}
\end{lemma}

Thanks to this result, we can now prove the next theorem;
indeed, in its formulation we include 
a source term and nonhomogeneous boundary conditions,
so it is slightly more general than the problem of the stiff string
we are studying.
\begin{theorem}
There exists at most one solution $u\in \cp{4}([0,L]\times[0, \infty))$
of the problem
\begin{equation}
\label{eq: cauchyProblemStiffNavier}
\begin{cases}
\dfrac{\partial^2 u}{\partial t^2} = c^2\dfrac{\partial^2 u}{\partial x^2} -
M^2 \dfrac{\partial^4 u}{\partial x^4} + f(x,t) & x\in(0,L),\ t>0\\
u(0,t) = u(L,t) = \delta (t) & t\geq0 \\
\partial_x^2 u(0,t) = \partial_x^2 u(L,t) = \mu (t) & t\geq0 \\
u(x,0) = \phi(x) & x\in[0,L] \\
\partial_t u (x,0)= \psi(x)& x\in[0,L]\, .
\end{cases}
\end{equation}
This is also true if, instead of conditions on $\partial_x^2 u$,
we put conditions on $\partial_x u$:
$\partial_x u(0,t) = \partial_x u(L,t) = \eta (t)$.

\begin{proof}
Let $u_1$ and $u_2$ be two solutions of the problem 
\eqref{eq: cauchyProblemStiffNavier}.  
Since the PDE is linear, 
$u := u_1 - u_2$ 
solves the homogeneous problem
\begin{equation}
\begin{cases}
\dfrac{\partial^2 u}{\partial t^2} = c^2\dfrac{\partial^2 u}{\partial x^2} -
M^2 \dfrac{\partial^4 u}{\partial x^4} & x\in(0,L),\ t>0\\
u(0,t) = u(L,t) = 0& t\geq0 \\
\partial_x^2 u(0,t) = \partial_x^2 u(L,t) = 0 & t\geq0 \\
u(x,0) = 0 & x\in[0,L] \\
\partial_t u (x,0) = 0& x\in[0,L]\,.
\end{cases}
\end{equation}
By lemma \ref{lemma: energy}, 
the non negative quantity 
$\mathcal{E}(u)$ 
is constant in~$t$. 
But at time $t=0$,
\begin{align*}
\mathcal{E}(u) \big|_{t = 0} 
&= 
\dfrac{1}{2} \int_0^L 
\!
\bigg(
(\partial_t u)^2 + c^2(\partial_x u)^2 + M^2(\partial_x^2 u)^2 
\bigg)
\bigg|_{t = 0}
\!\!\!\!
\d x 
= 0\,,
\end{align*}
so $\mathcal{E}(u) \equiv 0$ for all time, and thus
\begin{equation}
(\partial_t u)^2 + c^2(\partial_x u)^2 + M^2(\partial_x^2 u)^2 = 0 
\quad \Longrightarrow \quad
\partial_t u = 0 \,,  \   
\partial_x u = 0\,  \   
(\text{and }
\partial_x^2 u = 0)\,.
\end{equation}
Since all partial derivatives of first order of $u$ are zero, 
$u$ is a constant function. 
Finally, at $t=0$, $u=0$, so $u \equiv 0$ for all time $t\geq 0$. 
Therefore, $u_1 = u_2$.

The proof for boundary conditions on $\partial_x u$ is completely analogous.
\end{proof}
\end{theorem}

\section{Scales, spectrum and dissonance curves}

\label{sec:dissonance}

As we mentioned in the introduction,
if a spectrum is given,
by analysing its dissonance curve,
one can try to find appropriate scales for it.
In this section we provide some details for this analysis
following 
\cite{Sethares}.

In 1965,
R.~Plomp and W.J.M.~Levelt
\cite{PlompLevelt}
performed an experiment 
aiming to measure the dissonance of two pure tones 
in terms of their distance;
this dissonance was evaluated by many subjects,
and as a result
they concluded that the maximal degree of dissonance is attained
at roughly 1/4 of the critical bandwidth,
a concept from psychoacoustics 
that had been introduced and studied some years before
(see also
\cite{Roederer}).
Except for low frequencies, the width of a critical band
corresponds to an interval around a minor third.

Plomp and Levelt claimed that this could be extrapolated to complex tones,
so that the dissonance of a sound could be computed as the sum
of the dissonances of all the pairs of its partials.
More specifically,
by
taking a \emph{harmonic spectrum} with 6 partials (and equal loudnesses),
they obtained a dissonance curve similar to Helmholtz's,
showing points of local minimal dissonance for 
frequency ratios $\alpha$ equal to
1:1, 2:1, 3:2, 5:3, 4:3, 6:5, 5:4, 
and a maximum of dissonance near the semitone interval.
Figure~1 shows some of Plomp and Levelt's results.

\begin{figure}[H]
\vspace{-4mm}
\centering
\hfill
\includegraphics[width=0.30\linewidth]{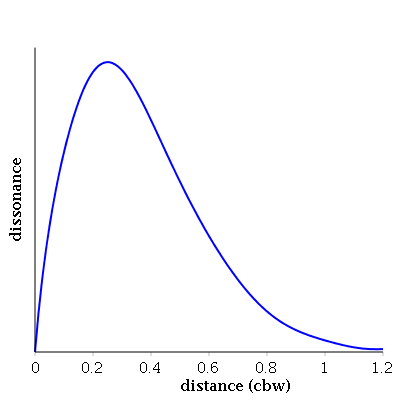}
\hfill
\includegraphics[width=0.35\linewidth]{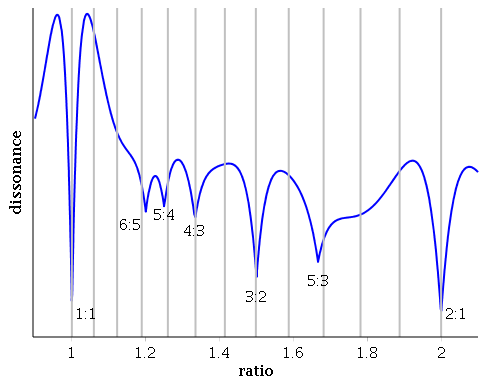}
\hfill\null
\caption{\small Graphics of Plomp and Levelt's results
\cite{PlompLevelt}.
(a) Dissonance of two pure tones as a function of their
distance measured in critical bandwidths;
extrapolated from experimental data.
(b) Theoretical model of the dissonance of two harmonic tones as a function of their
frequency ratio.
(The vertical lines mark the steps of the 12-edo scale.)}
\end{figure} 

More recently,
W.A.~Sethares
\cite{Sethares}
did a systematic study of the dissonance curve of 
several spectra,
and showed the close relationship between spectrum and scales;
his work includes the synthesis of artificial spectra
adapted to play music in exotic scales,
while still retaining some degree of consonance.

We want to apply this procedure to the spectrum of the string
with stiffness.
For this we need a specific expression of a function
modelling the dissonance.
Following
\cite{Sethares},
given two pure tones 
of frequencies $f_1 \leq f_2$ 
(expressed in Hz)
and loudnesses $\ell_1$, $\ell_2$
then the dissonance (in an arbitrary scale) 
can be expressed as
$
d(f_1,f_2,\ell_1,\ell_2) = 
\min(\ell_1,\ell_2) 
\left(
e^{-b_1 \, s \, (f_2-f_1)} - e^{-b_2 \, s \, (f_2-f_1)}
\right)
$,
where
$
s =
\frac{x^*}{s_1 \, f_1 + s_2}
$,
and the parameters are
$b_1 = 3.5$,
$b_2 = 5.7$,
$x^* = 0.24$, 
$s_1 = 0.021$
and
$s_2 = 19$.
The graph of this function reproduces the shape obtained by Plomp and Levelt,
Figure~1\,(a);
dissonance is measured in an arbitrary scale,
therefore usually 
we will normalize its expression so that it takes values between 0 and~1.
The preceding expressions and numbers give just a possible model for
the dissonance of two tones;
other models 
(see for instance 
\cite{Benson})
can be used,
and qualitatively the results are the same.

Then, 
if $\calF$ is a spectrum with 
frequencies $f_1 < \ldots < f_n$ and
loudnesses $\ell_1, \ldots, \ell_n$,
the dissonance of~$\calF$ is defined as the sum
of the dissonances of all the pairs of partials,
$
d_\calF = \sum_{i<j} d(f_i,f_j,\ell_i,\ell_j)
$.
Finally, the dissonance function of a given spectrum~$\calF$
is the function
that yields the dissonance of two tones
as a function of the ratio~$\alpha$
of their fundamental frequencies:
$
D_\calF(\alpha) = d_{\calF \cup \alpha \calF}
$,
where we denote by
$\alpha \calF$
the spectrum~$\calF$ with its frequencies scaled by the factor~$\alpha$,
and by
$\calF \cup \alpha \calF$
the union of both spectra.

The graph of the function $D_\calF$ is the \emph{dissonance curve}
of the given spectrum, and its analysis 
can help us to find an appropriate scale 
(and conversely: given an arbitrary scale,
is there an appropriate spectrum for it?).
Nevertheless, 
this is not so immediate,
and these results do not tell us how to construct a
scale.
For instance, 
consider the harmonic spectrum and its dissonance curve
as in Figure~1\,(b).
From a reference note
---a C, say---
one can form a just scale by adding other notes 
coinciding with the local minima of the dissonance curve:
G (3:2), F (4:3), A (5:3), E (5:4), etc. 
Notice, however, that from each of these new notes
we should consider again the dissonance curve 
in relation with the notes already chosen.
This analysis is simpler when we use an equal-step scale,
like 12-edo,
because the relative positions of the notes are the same.
In the same figure we see the abscissas of the 12-edo scale;
it is clear that
local minima of dissonance are attained near points
that are at a distance of
3, 4, 5, 7, 9 and 12 steps 
from any given note.


\begin{figure}[H]
\centering
\hfill
\includegraphics[width=0.6\linewidth]{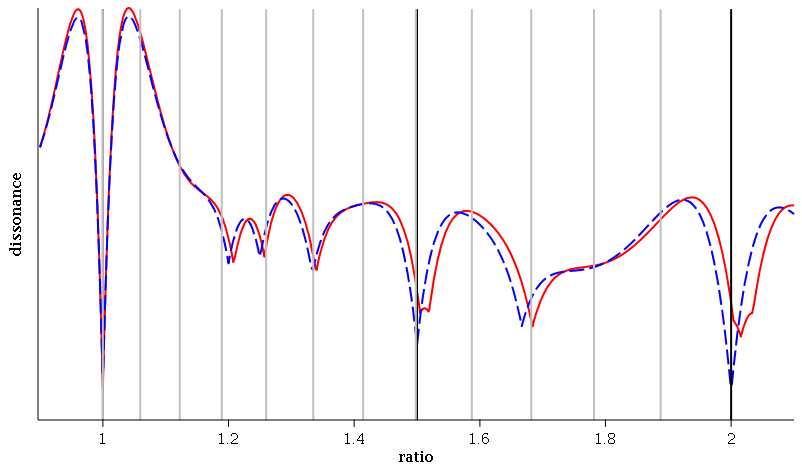}
\hfill\null
\caption{\small 
Comparison betwwen the dissonance curves of the harmonic spectrum
(dashed) and the stretched spectrum (solid) of a string with stiffness.
The grey vertical lines show the steps of the usual 12-edo scale;
the black ones show the just fifth and octave.
(We have used $B=0.0013$ for the sake of clarity.)}
\end{figure} 

Now let us apply this procedure to the piano.
As we have already noted in the preceding sections,
its strings have a certain degree of stiffness,
and, according to 
\eqref{eq: freq stiffness},
their spectrum is given by
$
f_n = n \,f_\circ \sqrt{1 + B n^ 2}
$,
for $n \geq 1$.
We can draw its dissonance curve
and we observe that, 
for small $B>0$,
the local minima of dissonance 
are slightly shifted to the right
with respect to those of the harmonic spectrum,
see Figure~2.

Notice in particular that
the octave and the fifth
(the most important intervals of Western music)
of the usual 12-edo scale
are noticiably flatter than 
the ``optimal'' octave and fifth
deduced from the stretched spectrum,
i.e.,
the corresponding intervals where this spectrum has a local minimum
of dissonance.
Therefore 
the 12-edo scale seems not to be the best choice to play music
as far as dissonance is concerned. 
This fact makes us wonder which is the ``best'' tuning for the piano, 
i.e., 
a tuning that fits better with the minima of the dissonance curve.
We give an answer to this question in the next section.

\section{Proposals for the piano tuning}
\label{sec:proposals}

We have just seen that,
due to the stiffness of the strings,
the spectrum of the piano is slightly stretched,
and therefore the minima of the dissonance curve 
does not coincide with the notes of the usual 
12-edo scale,
not to say 
other tunings like just intonation. 
Now our goal is to find a scale that preserves,
as much as possible,
the consonance of the main intervals of music.
We will restrict our search to scales with equal steps,
because we want to preserve the freedom 
to modulate to arbitrary tonalities
---this is especially important for piano music.
So, if $r$ is the frequency ratio of the step of the scale,
and $f$ is the frequency of its fundamental note,
the frequencies of all the notes are
$f, r \,f, r^2 f, r^3f, \ldots$

We will follow two procedures.
The first one is based on the 
\emph{coincidence of a couple of partials}:
then their beats disappear 
and we avoid their dissonance, 
as it is explained in the Appendix.
We will explore three possible choices for the step
and see what do they imply for the dissonance curve.
The second one is to define an 
\emph{average dissonance} 
as a function of the step and try to minimize it.

It should be remarked that in this study we assume that the
stiffness parameter $B$ is the same for all the strings.
This is approximately true in the middle third of the keyboard
\cite{Fletcher}. 
For the lower third of the keyboard,
the stiffness parameter of the strings is very low,
indeed they are manufactured in a special way,
so that a different analysis would be required;
besides this, 
the overall dissonance in this region is high.
For the upper third of the keyboard,
the upper partials are weak 
(and even become rapidly inaudible),
so that their effect on the dissonance can be neglected.

In all calculations we will use the obtained formula for the partials, 
$f_n = n \,f_\circ \,\sqrt{1 {+} B n^ 2}$
($n \geq 1$),
as well as the expresions of the dissonance functions 
defined in Section~3. 
We will consider $B \in [0.0004, 0.002]$ 
\cite{Fletcher}, 
but we will also see that we recover the results for the harmonic case when 
$B \to 0$.

\subsection{Coincidence of partials}

Here our strategy to construct a scale \emph{close to 12-edo} is as follows.

\begin{itemize}
\itemsep 0pt
\item 
We consider a fixed note of fundamental frequency $f_1$. 
Suppose we have already fixed a second note,
of fundamental frequency~$\overline{f}_1$.
Then we divide the interval
$\overline{f}_1:f_1$ in $p$ equal parts,
thus obtaining a step whose frequency ratio is
$r = (\overline{f}_1/f_1)^{1/p}$.
We will choose the number of parts~$p$
that makes
the step~$r$ 
to be the closest possible
to the frequency ratio of the 12-edo semitone,~$2^{1/12}$. 
\item
So we have to properly choose the second note~$\overline{f}_1$.
We base this choice upon the spectrum of the notes,
$(f_i)_{i \geq 1}$ and
$(\overline{f}_i)_{i \geq 1}$.
In our particular case,
we seek the coincidence of some partials.
So, we define the tuning
$\mathcal{A}_{m,n}$
as the one obtained by letting 
the $m$-th 
partial $f_m$ of the first note to coincide with 
the $n$-th partial~$\overline f_n$ of the second note. 
Equating $f_m = \overline{f}_n$ 
determines $\overline{f}_n$ and therefore $\overline{f}_1$.
\item
Finally, 
if $r = r_{m,n}$ is the step of
$\mathcal{A}_{m,n}$,
the notes of the scale are
$r^k f_1$, for integer values of~$k$.
\end{itemize}

We have noticed before that 
the coincidence of some partials does not necessarily imply consonance. 
However, this analysis is meaningful because it can be directly applied
to actual tuning,
since it is easy to tune an interval by letting beats disappear;
moreover, we will see later that one of our proposals
will be especially good in terms of dissonance.

As the octave and the fifth are the most important intervals in music, 
three natural tunings can be considered:
\begin{itemize}
\itemsep 0pt
\item $\mathcal{A}_{2,1}$: 
the second partial $f_2$ of the first note coincides with 
the first partial~$\overline f_1$ of the second note. 
(We try to minimize the beats of the octave.)
\item $\mathcal{A}_{3,1}$: 
the third partial $f_3$ of the first note coincides with 
the first partial~$\overline f_1$ of the second note. 
(We try to minimize the beats of the twelfth.)
\item $\mathcal{A}_{3,2}$: 
the third partial $f_3$ of the first note coincides with 
the second partial~$\overline f_2$ of the second note. 
(We try to minimize the beats of the fifth.)
\end{itemize}

In the following figure we show
a schematic description of these tunings:

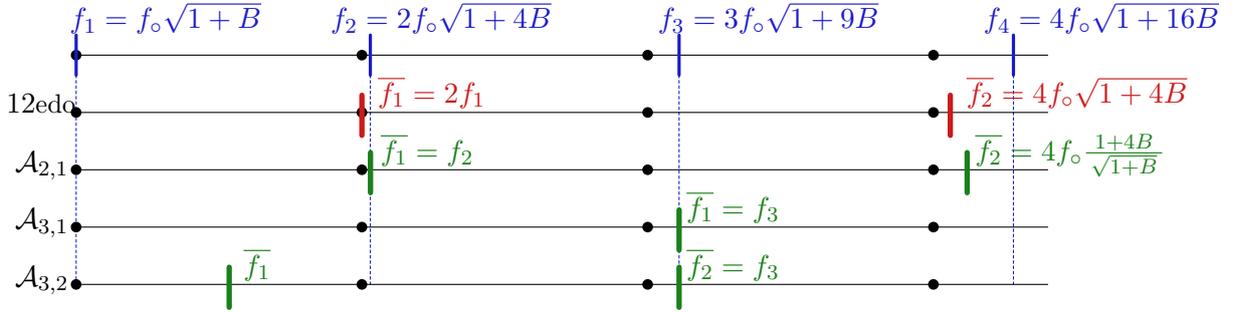
\begin{figure}[H]
\definecolor{red}{rgb}{0.8,0.1,0.1}
\definecolor{darkgreen}{rgb}{0.1,0.5,0.1}
\definecolor{blue}{rgb}{0.1,0.1,0.8}

\begin{tikzpicture}%
[line cap=round,line join=round,>=triangle 45,x=3.76cm,y=3.8cm]
\clip(-0.235,-0.9) rectangle (4.1,0.2);


\draw (0,0.)-- (3.4,0.); 

\fill (0,0) circle (2pt);
\fill (1,0) circle (2pt);
\fill (2,0) circle (2pt);
\fill (3,0) circle (2pt);

\draw [line width=1.2pt,color=blue] (0,0.07)-- (0,-0.07);
\draw [line width=1.2pt,color=blue] (1.0295,0.07)-- (1.0295,-0.07);
\draw [line width=1.2pt,color=blue] (2.11,0.07)-- (2.11,-0.07);
\draw [line width=1.2pt,color=blue] (3.28,0.07)-- (3.28,-0.07);

\draw [color=blue] (-0.058,0.22) node[anchor=north west]  {$f_1 = f_\circ\sqrt{1+B}$};
\draw [color=blue] (0.855,0.22) node[anchor=north west]  {$f_2 = 2f_\circ\sqrt{1+4B}$};
\draw [color=blue] (2.0,0.22) node[anchor=north west] {$f_3 = 3f_\circ\sqrt{1+9B}$};
\draw [color=blue] (3.14,0.22) node[anchor=north west]  {$f_4 =4 f_\circ\sqrt{1+16B}$};

\draw [color=blue][dash pattern=on 1pt off 1pt] (0,-0.07)--(0,-0.8);
\draw [color=blue][dash pattern=on 1pt off 1pt] (1.0295,-0.07)--(1.0295,-0.8);
\draw [color=blue][dash pattern=on 1pt off 1pt] (2.11,-0.07)--(2.11,-0.8);
\draw [color=blue][dash pattern=on 1pt off 1pt] (3.28,-0.07)--(3.28,-0.8);

\draw (-0.28,-0.1) node[anchor=north west]   {12\small edo};
\draw (-0.,-0.2)-- (3.4,-0.2);

\fill (0,-0.2) circle (2pt);
\fill (1,-0.2) circle (2pt);
\fill (2,-0.2) circle (2pt);
\fill (3,-0.2) circle (2pt);

\draw [line width=2.pt,color=red] (1.,-0.14)-- (1.,-0.28);
\draw [line width=2.pt,color=red] (3.059,-0.14)-- (3.059,-0.28);

\draw [color=red](1.02,-0.04) node[anchor=north west] 
{$\overline{f_1}= 2f_1 $};
\draw [color=red](3.08,-0.04) node[anchor=north west]  
{$\overline{f_2} = 4f_\circ\sqrt{1+4B}$};

\draw (-0.25,-0.3) node[anchor=north west]  
{$\mathcal{A}_{2,1}$};
\draw (-0.,-0.4)-- (3.4,-0.4);

\fill (0,-0.4) circle (2pt);
\fill (1,-0.4) circle (2pt);
\fill (2,-0.4) circle (2pt);
\fill (3,-0.4) circle (2pt);

\draw [line width=2pt,color=darkgreen] (1.03,-0.34)-- (1.03,-0.48);
\draw [line width=2pt,color=darkgreen] (3.118,-0.34)-- (3.118,-0.48);

\draw [color=darkgreen](1.03,-0.24)   node[anchor=north west] 
{$\overline{f_1}= f_2$};
\draw [color=darkgreen](3.11,-0.24) node[anchor=north west]  
{$\overline{f_2}= 4f_\circ \frac{1+4B}{\:\sqrt{1+B}\:}$};

\draw (-0.25,-0.5) node[anchor=north west]  {$\mathcal{A}_{3,1}$};
\draw (-0.,-0.6)-- (3.4,-0.6);

\fill (0,-0.6) circle (2pt);
\fill (1,-0.6) circle (2pt);
\fill (2,-0.6) circle (2pt);
\fill (3,-0.6) circle (2pt);

\draw [line width=2.pt,color=darkgreen] (2.11,-0.54)-- (2.11,-0.68);

\draw [color=darkgreen](2.1,-0.44) node[anchor=north west] {$\overline{f_1} = f_3$};

\draw (-0.25,-0.7) node[anchor=north west]  {$\mathcal{A}_{3,2}$};
\draw (-0.,-0.8)-- (3.4,-0.8);

\fill (0,-0.8) circle (2pt);
\fill (1,-0.8) circle (2pt);
\fill (2,-0.8) circle (2pt);
\fill (3,-0.8) circle (2pt);

\draw [line width=2.pt,color=darkgreen] (0.5356,-0.74)-- (0.5356,-0.88);
\draw [line width=2pt,color=darkgreen] (2.11,-0.74)-- (2.11,-0.88);

\draw [color=darkgreen](0.55,-0.64) node[anchor=north west] {$\overline{f_1}$};	
\draw [color=darkgreen](2.1,-0.64) node[anchor=north west] {$\overline{f_2} = f_3$};

\end{tikzpicture}

\caption{\small
The partials of the stretched spectrum are represented in the upper line;
they are compared with the harmonic ones (black nodes).
In the lower lines the partials of four tunings are shown:
the usual just octave (12-edo), 
the octave of $\mathcal{A}_{2,1}$, 
the twelfth of $\mathcal{A}_{3,1}$, and 
the fifth of $\mathcal{A}_{3,2}$.}
\label{fig: distribution-partials}
\end{figure}

Once we have tuned our interval
$\overline{f}_1 : f_1$,
we divide it in $p$ equal parts:
12 for the octave, 19 for the twelfth, and 7 for the fifth.
These steps are the semitones of the corresponding tuning.
Their frequency ratio is given by
$(\overline{f_1}/f_1)^{1/p}$; 
in our cases, they are:
\begin{equation}
r_{2,1} = 2^{1/12} \left ( \dfrac{1+4B}{1+B} \right )^{1/24}, \ \ \ \ 
r_{3,1} 
= 3^{1/19} \left(\dfrac{1+9B}{1+B}\right)^{1/38}, \ \ \ \ 
r_{3,2} 
=  \left( \dfrac{3}{2}\right)^{1/7}\left(  \dfrac{1 + 9B}{1 + 4B}\right)^{1/14}.
\end{equation}

Now, a given choice of the step~$r$ defines a scale, 
and we can locate its notes in the dissonance curve.
The idea is that,
by stretching the gap between the notes
(vertical lines in Figure~2),
possibly the new notes will fit better the minima of the dissonance curve
of the stretched spectrum
(solid line in the figure).

One of our main goals is to tune the octave, 
the most important interval in music. 
Therefore, 
we analyze in particular how the new octaves generated by these steps 
(the ratio obtained by $r^{12}$, for each of the chosen values of~$r$) 
fit the minimum of the dissonance curve near the frequency ratio~2.
The same can be done with the fifth by analyzing the ratios $r^7$
near the ratio 3:2 in the dissonance curve.
The results are shown in Figure~4.

\begin{figure}[H]
\includegraphics[width=0.5\linewidth, height=0.25\linewidth]{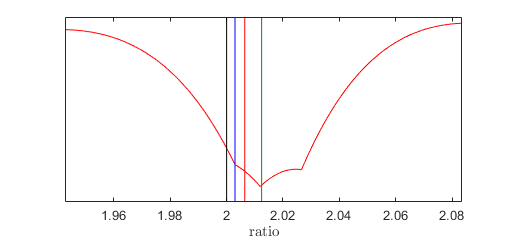}%
\includegraphics[width=0.5\linewidth, height=0.25\linewidth]{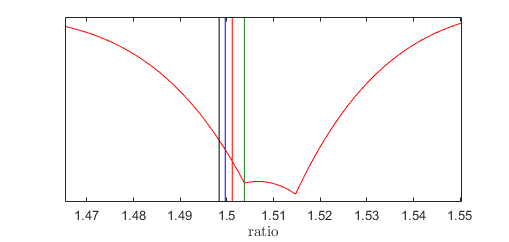}%
\caption{\small
(a)~%
The dissonance curve of the stretched spectrum
($B=0.001$)
near the just octave;
vertical lines 
show the octave generated by different steps~$r$:
from left to right, 
$r= 2^{1/12}$ (12-edo),
$r=r_{2,1}$,
$r=r_{3,1}$,
and
$r=r_{3,2}$.
(b)~%
The same near the just fifth.
}
\label{fig:octaves}
\end{figure} 

From that figure
it appears that the tuning
$\mathcal{A}_{3,2}$ fits better than the others
the minimum of dissonance at the octave
and also at the fifth. 
For the octave this may seem paradoxical because $\mathcal{A}_{2,1}$ 
was set to tune the octave ad hoc, 
but actually this tuning only makes the dissonance caused 
by a single pair of partials to disappear,
whereas other partials may give rise to higher dissonance. 
This suggests also that our study should consider all the other intervals,
because we are not controlling their dissonance.
In the next section we make a proposal to deal with this.


\subsection{Minimization of dissonance}

The preceding analysis can be completed by performing a general
study of the dissonance of all the intervals of the scale.
We would like to find the semitone $r$ which minimizes (in some sense) 
the total dissonance of the scale.

In the most general setting, 
we could define the mean dissonance of a scale
as a weighted sum of the dissonances of all couples of notes.
The weighting is necessary because 
not all intervals are equally used in music,
and different intervals have different musical roles;
therefore their consonances are not equally important.

For an equal step scale it is enough to consider the dissonances of
all the notes with respect to a given one,
that is,
the dissonances between the fundamental note of the scale
(with frequency~$f_1$) 
and the others 
(with frequencies~$r^k f_1$).
More specifically,
given a semitone $r$, 
we define the \emph{mean dissonance} of the 
equal step scale generated by~$r$ as
a weighted average
of the dissonances of all the
intervals from the fundamental note of the scale 
within the range of an octave,
that is:
\begin{equation}
D_{\mathrm{m}}(r) := 
\sum_{k = 1}^{12} w_k \,
D_\calF( r^k)\,,
\end{equation}
where $\mathcal{F}$ is the spectrum of the fundamental note~$f_1$, 
including frequencies and loudnesses,
and 
$w = (w_k)$ is a vector of weights.
In order to give preeminence to the octave, the fifth, etc,
in the following calculations we have used 
$w = ( 1, 1, 4, 4, 5, 2, 6, 4, 4, 2, 1, 10)$.
Intervals larger than an octave could be considered in the sum;
we omit them because their effect on the dissonance is small
and we have to cut the sum at some point.

If we minimize numerically the function $D_{\mathrm{m}}(r)$ 
on the interval $r\in[1.0585, 1.061]$,
near the 12-edo semitone $2^{1/12}$, 
we find different values for the minimum point $r^*$ depending on~$B$. 
The results are shown in Figure~5~(a).
\begin{figure}[H]
\includegraphics[width=0.45\linewidth]{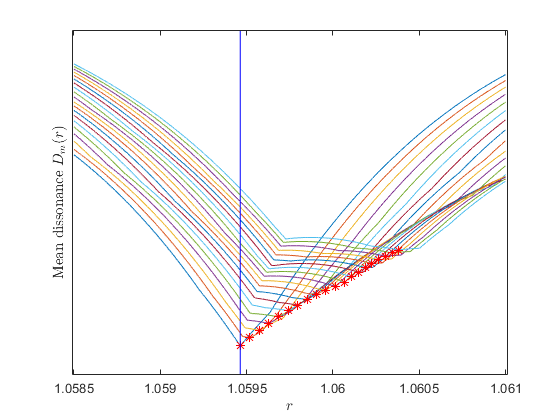}
\hfill
\includegraphics[width=0.45\linewidth]{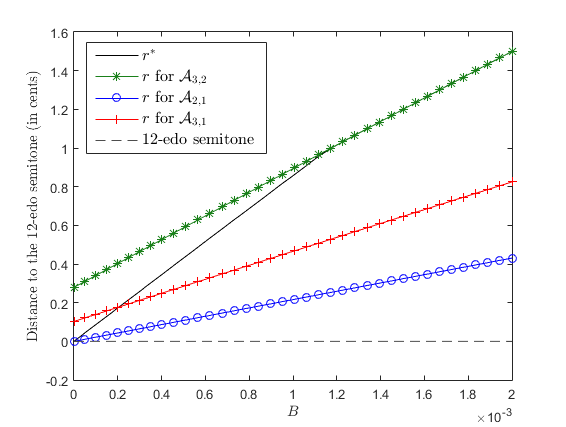}
\caption{\small
(a) Plot of $D_{\mathrm{m}}(r)$ for different values of $B \in [0,0.002]$. 
For each~$B$ the marked point corresponds to the minimum~$r^*$. 
The vertical line represents the 12-edo semitone. 
(b) Distance to the 12-edo semitone of the three tunings and 
the optimal semitone (in cents).}
\end{figure} 

We want to compare this optimal semitone
with the semitones of the three tuning proposals considered before.
For each~$B$,
we compute the distance of these four semitones to the 12-edo semitone;
this is shown in Figure~5~(b).
As we can see, 
the semitone of $\mathcal{A}_{3,2}$ 
approximates the optimal semitone:
(i) better than the 12-edo semitone if $B$ is higher than 0.00025
(ii) better than the other proposals if $B$ is higher than 0.0005,
and
(iii) coincides with the optimal semitone if $B$ is higher than 0.001.
Notice also that the optimal semitone coincides with the 12-edo semitone
when $B=0$.

The graphics in Figure~5
have been computed with Matlab. 
We have used a spectrum of 6~partials,
the value $f_\circ = 440$~Hz, 
the weighting vector as defined before,
and the loudnesses inversely proportional to the number of the partial.
Nevertheless, 
we have also checked that the results obtained are quite similar
if we use the same loudness for all partials,  
the same weights for all intervals, 
or 7 partials instead of~6.

\section{Conclusions}

We have studied the spectrum of strings with stiffness
modelled according to the Euler--Bernouilli model.
Although this was already known, 
we have done a mathematically rigorous derivation of it
using elementary techniques.
We have applied this result and the theory of dissonance
to study the tuning of the piano with a scale of equal steps.
We have followed two approaches: 
one is to define a scale based on
the coincidence of some specific partials;
the other one is to define an 
average dissonance of a scale
and trying to minimize it 
as a function of the step.
It appears that a good solution 
is to tune a note and its fifth  
by forcing their 3rd and 2nd partials, respectively, 
to coincide.

\subsection*{Acknoledgments}

X.G. acknowledges partial financial support by
Generalitat de Catalunya project 2014--SGR--634.
The authors wish to thank the anonymous referee whose comments
have greatly improved the paper.

\appendix
\section*{Appendix: sound and music}

In this appendix we summarize some basic information
about sound and music.
This can be found in many books, as for instance
\cite{Benson,RossingMooreWheeler}.

\def\paragraph#1{\smallskip\noindent\textbf{#1}\quad\ignorespaces}
\paragraph{Sound, pitch, spectrum}
\emph{Sound} is both an oscillation of the air pressure,
and also the auditory sensation it creates.
Besides duration, 
sound has three main perceptive attributes:
loudness, pitch and timbre.
These are related to physical attributes:
amplitude, frequency and spectrum.
However, these relations are by no means simple.

Let us consider the \emph{pitch}, 
a quality that allows sounds to be ordered
from lower to higher pitches.
A \textit{pure tone} of frequency~$\nu$ and amplitude~$A$
is described by a sinusoid
$A \sin(2\pi \nu t)$,
and its pitch can be identified with the frequency.
A musical sound 
is usually a superposition of pure tones 
(the \textit{partials})
of several frequencies and amplitudes;
these constitute the \emph{spectrum} of the sound.
For instance,
most wind and string instruments have 
\emph{harmonic spectrum},
i.e.,
their spectral frequencies are integer multiples
of a 
\emph{fundamental frequency}~$f_1$,
that is,
$f_n = n\,f_1$,
with $n \geq 1$.
Such a sound is perceived to have a pitch identified with frequency~$f_1$.
However, not every sound can be attributed a pitch;
some musical instruments, for instance most drums, have indefinite pitch.

\paragraph{Intervals, octave, semitone, cents}
The difference between two pitches
is called \emph{interval}. 
The pitch perception obeys two fundamental rules.
One is the logarithmic correspondence:
the interval from two pitches of frequencies $\nu_1, \nu_2$ 
only depends on their \emph{frequency ratio} $\nu_2 {\,:\,} \nu_1$. 
The other one is the octave equivalence:
two pitches an \emph{octave} apart (ratio 2:1) 
are musically equivalent.

One can measure intervals in the multiplicative scale by their frequency ratio, 
or in the additive scale by their size expressed in octaves, for instance.
A frequency ratio of $r$ corresponds to $\log_2 r$ octaves.
Other important intervals are the 
\emph{semitone},
which is 1/12 of an octave
(therefore its ratio is $2^{1/12}$),
and the \emph{cent},
which is 1/100 of a semitone.

The human ear is exceedingly sensitive to pitch perception.
The \emph{difference limen} (or just noticeable difference) 
between two tones can be,
depending on the frequency and intensity,
as small as 10 cents.
It can be much smaller if both sounds are played together.
Therefore it is of the greater importance to correctly tune
a musical instrument.

\paragraph{Notes, scales, 12-edo}
In some instruments 
(e.g.\ the violin) 
the player can play virtually any pitch 
within its playing range.
This is not true for other instruments
(e.g.\ the piano, or most wind instruments), 
where only a finite set of pitches is directly playable.
A selection of pitches to play music is called a
\emph{scale},
and its elements are the 
\emph{notes} of the scale.
The construction of these scales 
is one of the fundamental problems in music theory.
Notice that if we have chosen a scale on a theoretical basis,
then we have to adjust or to \emph{tune} the pitches of the notes of the instrument 
to the pitches of the scale;
therefore one frequently says \emph{tuning system} to mean a scale.

From ancient times it is known that
two similar strings sounding together
are more pleasant when their 
fundamental frequencies
are in a ratio of \textit{small integers}.
These intervals are called \emph{just},
and, in addition to the octave, the most important ones are 
the fifth (ratio 3:2),
the fourth (4:3)
and the major and minor thirds (5:4 and 6:5).
(These names have a historic origin, of course.)
So one would look for scales
whose notes define such intervals.
But, of course, other intervals will appear, 
and maybe they will be not so pleasant.
Moreover, the evolution of the musical language 
during the last centuries has added more requirements to the scales,
and as a result
the problem of defining a scale 
does not have a universal optimal solution.
What is more,
from antiquity to modern times, 
\emph{dozens} of scales have been proposed and put into practice
\cite{Grove}.

Among all of these scales,
there is one that is pervasive in Western music since 19th century.
It is the so-called equal temperament,
and consists of 12 equal divisions of the octave (12-edo).
The explanation for this choice is that 
the 12-edo scale yields an excellent approximation of the just fifth
($2^{7/12} \approx 3{\,:\,}2$) 
and the just fourth,
but also acceptable approximations of the just thirds.

It is worth noting that, for instance, 
the same name ``fifth'' is applied to two intervals
that are indeed different:
the just fifth and the 12-edo fifth. 
This is usual:
the traditional name of an interval
applies to all the intervals that have the same musical function
regardless of their exact tuning.
The same happens with the notes: 
A4 has nowadays a ``standard pitch'' of 440\,Hz,
but it is usual to tune this note to 442\,Hz, for instance.
In past times its values were much more diverse.

Due to the logarithmic correspondence,
from the viewpoint of music theory,
to define a scale 
one can fix the pitch $\nu_0$ of a
\emph{fundamental note} 
with some degree of arbitrariness;
what is really important are
the intervals $r_k = \nu_k {\,:\,} \nu_0$ 
between this note and the other ones, $\nu_k$.
From these intervals,
and the fundamental note,
the other notes can be reconstructed as
$\nu_k = r_k \,\nu_0$.
Alternatively, 
one can define a scale by giving the 
\emph{steps} between consecutive notes,
$\nu_k {\,:\,} \nu_{k-1}$; 
for instance, 
the 12-edo scale has equal steps of ratio $2^{1/12}$.

\paragraph{Beats, dissonance, consonance}
Using trigonometric identities
it is easily proved that the superposition of two pure tones
$A \sin(2\pi \nu_1 t)$
and
$A \sin(2\pi \nu_2 t)$
can be expressed as
\[
2\, A
\cos \left( 2\pi \frac{\nu_1-\nu_2}{2} t\right) 
\sin \left( 2\pi \frac{\nu_1+\nu_2}{2} t\right)
\,.
\]
If the frequency difference
$\nu_1{-}\nu_2$ 
is small
(about less than 10--15\,Hz),
this is perceived as a sound of frequency
$\nu = \frac{\nu_1+\nu_2}{2}$
with slowly fluctuating amplitude;
these are the \emph{beats}.
If the frequency difference is somewhat bigger,
one perceives some \textit{roughness}.
When the difference is even higher, 
then one perceives two separate tones
\cite[pp.\,37--40]{Roederer}.
This roughness gives rise to the notion of
\emph{sensory dissonance};
this is the only notion of dissonance
we are concerned about,
though there are others
\cite{PlompLevelt}
\cite[chap.\,5]{Sethares}.

Now, consider two (or more) complex tones sounding together:
they have many partials that may be close in frequencies. 
In the middle of 19th century
H.~Helmholtz
described the dissonance as the roughness produced by close partials,
and the \emph{consonance} as the exceptional condition
where this roughness almost disappears.
By computing the beats of the partials of two harmonic tones,
Helmholtz showed that
the aforementioned just intervals 
(octave, fifth, fourth, thirds)
are more consonant than others
\cite[p.\,193]{Helmholtz},
in good agreement with music theory.
This result can be easily understood:
just notice that,
if the fundamental frequencies $f, \tilde f$ of two harmonic tones
are in a ratio of small integers
$
\tilde{f} {\,:\,} f 
= 
\ell {\,:\,} k
$, 
then 
the $\ell$-th partial of the first tone
will be coincide with 
the $k$-th partial of the second one;
a slight change of~$\tilde{f}$
would lead to close but different partials,
and therefore to some roughness.


\end{document}